\documentstyle[multicol,aps,prb,epsfig,amstex]{revtex}

\tighten

\begin{document}
                                      
\draft
\title{High-Frequency Impedance of Driven Superlattices}
\author{A.-K. Jappsen$^{a}$, A. Amann$^{a}$, A. Wacker$^{a}$, 
E. Schomburg$^{b}$, E. Sch{\"o}ll$^{a}$}
\address{$^{a}$Institute for Theoretical Physics, Technical University of Berlin, Hardenbergstr.~36, 
D-10623~Berlin, Germany\\
$^{b}$ Institute for Applied Physics,
Universit{\"a}t Regensburg,
D-93040 Regensburg, Germany}

\date{\today}
\maketitle
\begin{abstract}
The complex impedance of a semiconductor superlattice biased into the regime
of negative differential conductivity and driven by an additional GHz ac
voltage is computed. From a simulation of the nonlinear spatio-temporal
dynamics of traveling field domains we obtain strong variations of the
amplitude and phase of the impedance with increasing driving frequency.
These serve as fingerprints of the underlying quasiperiodic or frequency
locking behavior. An anomalous phase shift appears as a result of phase
synchronization of the traveling domains.
If the imaginary part of the impedance is compensated by 
an external inductor, both the frequency and the intensity of the
oscillations strongly increase.
\end{abstract}
 
\begin{multicols}{2} 
\narrowtext

Semiconductor superlattices (SL)
show pronounced negative differential conductivity
(NDC) \cite{ESA70}.
If the total bias is chosen such that the average electric field is in the
NDC region, stable inhomogeneous field distributions 
(field domains)\cite{ESA74}
or self-sustained oscillations \cite{KAS95} with frequencies up
to 150 GHz at room temperature\cite{SCH99h}
appear. Which of these scenarios occurs depends on 
bias, doping, temperature, and the properties of the injecting contact
\cite{WAC97a,PAT98,SAN99,WAN99e,STE00a,AMA01}. For a recent overview 
see Ref.~\cite{WAC02}. 

In order to apply the self-sustained oscillations as a high--frequency 
generator  in an electronic device,
it is crucial to know the response of the SL in an external circuit.
A key ingredient for the analysis is the complex impedance of the SL
in the respective frequency range. This is the subject of this paper
where the complex impedance is evaluated numerically by 
imposing an additional ac
bias to the SL. The interplay of the self-sustained oscillations
and the external frequency causes a variety of interesting phenomena such as
frequency locking, quasi-periodic and chaotic behavior which
has been extensively studied both theoretically 
\cite{BUL95,CAO99a,SAN01} and experimentally \cite{ZHA96,SCH02b}.
In contrast to those studies we focus on the response to the circuit
and concurrent phase synchronization phenomena
in this work.

We describe the dynamical evolution of the SL by rate
equations for the electron densities in the quantum wells,
together with Poisson's equation for the electric fields.
The current densities 
$J_{j\to j+1}$ between adjacent quantum wells
are evaluated within the model of sequential tunneling,
and Ohmic boundary conditions with a contact conductivity $\sigma$
are used. For details see Refs.~\cite{AMA01,WAC02}. 
Here we apply
a periodic bias signal
$U(t)=U_{\rm dc}+U_{\rm ac}\sin(2\pi\nu_1 t)$
and study the total current 
$I(t)=A\sum_{j=0}^NJ_{j\to j+1}/(N+1)$.
In particular we consider the
Fourier component $I_{\rm ac}(\nu_1) \sin(2\pi\nu_1t-\phi)$ which gives the
complex impedance
\begin{equation}
Z(\nu_1)=\frac{U_{\rm ac}}{I_{\rm ac}(\nu_1)}e^{i\phi}\, .
\label{EqZdef}
\end{equation}

As an example we consider the SL structure studied in 
Ref.~\cite{SCH02b}.
It consists of $N=120$ GaAs wells of width 4.9 nm 
separated by 1.3 nm AlAs barriers.
The sample is n-doped with a density of $8.7\times 10^{10}{\rm cm}^{-2}$
per period and the sample cross
section is $A=64\, (\mu {\rm m})^2$. 
We estimate the energy broadening
$\Gamma= 15$ meV, which effectively gives the sum of phonon, impurity 
and interface roughness scattering rates.  
All calculations are performed at room temperature.

\begin{figure}
\noindent\includegraphics[width=8.5cm]{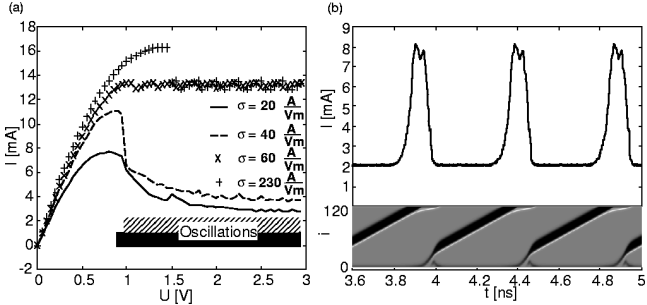}
\caption[a]{(a) Calculated current-voltage characteristics for different 
contact conductivities $\sigma$, and $U_{\rm ac}=0$.
For $\sigma=20$ A/Vm and 40 A/Vm self-sustained current oscillations
are found in the indicated bias ranges. In this case we display the
time-averaged current. 
(b) Self-sustained oscillations for $U_{\rm dc}=2$ V, $\sigma=20$ A/Vm.
The time series of the current $I(t)$ and the space-time plot of the electron
densities $n_i(t)$ in the quantum wells is shown.
Black indicates low electron density (depletion front), white indicates
high electron density (accumulation front).
The emitter is at the bottom, the collector at the top.
}
\label{FigKenn}
\end{figure}

In Fig.~\ref{FigKenn}a we display the calculated current--voltage 
characteristic for $U_{\rm ac}=0$. While for large values of the contact 
conductivity $\sigma$
stationary field domains occur, self-sustained current oscillations 
(where pairs of accumulation and depletion fronts travel through the sample)
are found for lower $\sigma$ (Fig.~\ref{FigKenn}b). This scenario is quite
general for moderately doped samples \cite{SAN99,WAC02}. 
In the following we use $\sigma =20$ A/Vm  giving best agreement 
with the measured current-voltage characteristic.

\begin{figure}
\noindent\includegraphics[width=6.cm]{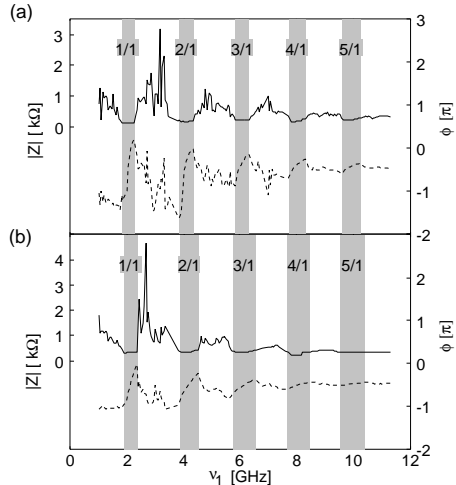}
\caption[a]{Amplitude $|Z|$ (full line) and phase $\phi$ (dashed line) of
the complex impedance $Z$ as a function of the driving
frequency for two different values of the ac bias: (a) $U_{ac}=0.2$V,
(b) $U_{ac}=0.6$V. The shaded areas indicate
locking intervals marked by $\nu_1/\nu_0^*$ ($U_{dc}=2$V, $\sigma=20$ A/Vm).
}
\label{FigZges}
\end{figure}

At $U_{\rm dc}=2$ V, the self-sustained oscillations exhibit a
frequency of $\nu_0=2.1$ GHz. Now we study the change in the  current
signal by imposing an additional ac-bias amplitude $U_{\rm ac}$ with
frequency $\nu_1$.  Fig.~\ref{FigZges} displays the absolute value and
the phase of the complex impedance given by Eq.~(\ref{EqZdef}). For
both strengths of the  ac amplitude  one observes strong variations
in the amplitude and phase of $Z$. In particular one observes
pronounced minima in $|Z|$ and a corresponding monotonic increase in
$\phi$ around frequencies $\nu_1$ which are approximately integer
multiples of $\nu_0$.  This is due to frequency locking \cite{SCH88a},
where the external frequency $\nu_1$ modifies the main oscillation
frequency to $\nu_0^*$ such that $\nu_1/\nu_0^*$ is a rational number.
While locking occurs in a rather wide frequency range for
$\nu_1/\nu_0^*\in \mathbb{N}$, locking into rational numbers with
larger denominators are less easy to observe (The corresponding widths
of the  locking intervals roughly follow the position in the Farey-tree
of the rational numbers, see \cite{SCH88a}. The width of the locking
ranges also increases with $U_{\rm ac}$.)  A more detailed examination
indicates that the local minima in  $|Z|$ correspond to the locking
regions, and $\phi$ increases with $\nu_1$ in
those intervals, while a
decrease is frequently observed outside these ranges where
quasi-periodic  or possibly chaotic behavior occurs.  Thus the
presence of several locking intervals\cite{SCH02b}  explains the variety
of peaks.

For high frequencies $\nu_1\gg \nu_0$ the variations of $Z$ become
less pronounced and $|Z|\sim 300\Omega$ and $\phi\approx -\pi/2$ is
observed. This corresponds to a capacitive current with an effective
frequency-dependent capacitance
$C_{\rm eff}(\nu_1)\approx 1/(2\pi \nu_1|Z| )$.  At $\nu_1=10$ GHz this
gives  $C_{\rm eff}\approx 50$ fF.
This capacitance  results from the interaction with the internal  front
dynamics in the SL structure; the intrinsic sample
capacitance  $\epsilon_r\epsilon_0A/(Nd)=10$ fF is significantly
smaller.

While we have only shown results for   $U_{\rm dc}=2$ V and
$\sigma=20$ A/Vm here,  we checked that the features discussed above
neither change for different biases nor  different contact
conductivities, albeit the main oscillation frequency $\nu_0$ and the
locking ranges change  slightly.

\begin{figure}
\noindent\includegraphics[width=8.5cm]{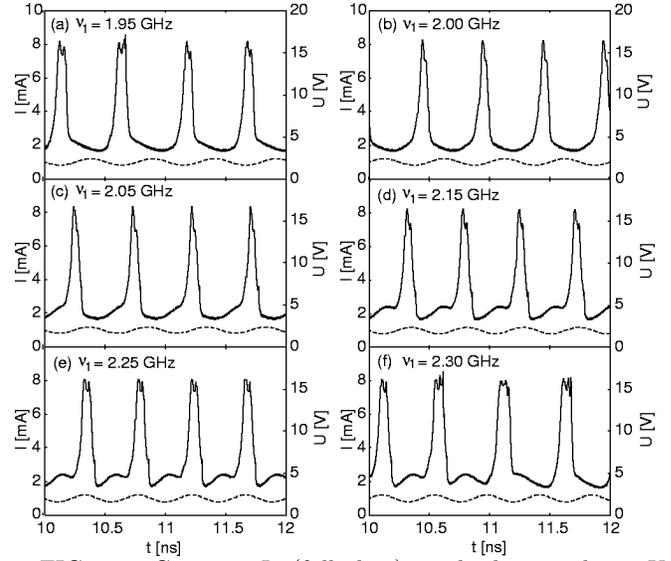}
\caption[a]{Current $I$ (full line) and driving bias $U$ (dashed line) versus
time for different driving frequencies $\nu_1$ ($U_{\rm ac}=0.4 V$,
$U_{dc}=2$V). 
}
\label{FigTimeresolved}
\end{figure}

Now we focus on the 1/1 locking region which is of interest if the
device is used as an oscillator. In Fig.~\ref{FigTimeresolved} we
show the current  and bias signal for different frequencies
$\nu_1$. In parts (a) ($\nu_1=1.95$ GHz)  and (f)  ($\nu_1=2.3$ GHz)
no complete locking occurs and the current  does not exhibit a
periodic signal.  Between these frequencies the current signal is
periodic with a frequency $\nu_0^*=\nu_1$  imposed by the external
bias. While the current peaks occur around the minima of the external
bias for $\nu_1=2$ GHz (corresponding to $\phi\approx -\pi$), the
delay between the current peaks and the bias maxima decreases with
frequency until they are approximately in phase at the upper boundary
of the locking range at  $\nu_1=2.25$ GHz (corresponding to
$\phi\approx 0$).  It is intriguing to note that this behavior is
opposite to the response of a damped linear oscillator, where the
phase between the driving signal and the response shifts from $0$ to
$\pi$
with increasing driving frequency. The current peaks correspond to the
formation of a domain at the emitter, while during the current minima the
domain traverses the sample. The domain transit velocity is larger for smaller
voltage. The phase shift between current and voltage adjusts such that the
domain velocity increases with increasing driving frequency during the whole
locking interval. This is a phase synchronization effect induced by the domain
dynamics.

\begin{figure}  
\noindent\includegraphics[width=8.5cm]{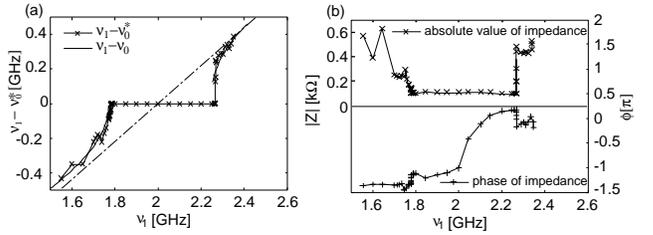}
\caption[a]{(a) Shift of the fundamental frequency $\nu_0^*$ of the 
current signal and (b) amplitude 
and phase of the complex impedance as a function of the driving frequency
$\nu_1$ in the vicinity
of the 1/1 locking region ($U_{\rm ac}=0.2$ V, $U_{dc}=2$V).
}
\label{FigDetail}
\end{figure}

Let us analyze the behavior close to the onset of the
locking interval in detail, see Fig.~\ref{FigDetail}.  
The main frequency component of the current signal is given by $\nu_0^*$.
In the locking
region it is equal to the frequency of the driving bias $\nu_1$ such
that $\nu_0^*-\nu_1$ vanishes in a finite range of $\nu_1$. In
contrast, far away from the locking $\nu_0^*$ is essentially given by
the free oscillation frequency $\nu_0$, thus $\nu_0^*-\nu_1$ exhibits a linear
relation.  Close to the boundaries of the locking region a square root
behavior $\nu_0^*-\nu_1\propto
\sqrt{|\nu_1-\nu_1^{\rm crit}|}$ can be detected, which is a general
feature of frequency locking, see, e.g.,  \cite{PEI92}. 

\begin{figure}  
\noindent\includegraphics[width=8.5cm]{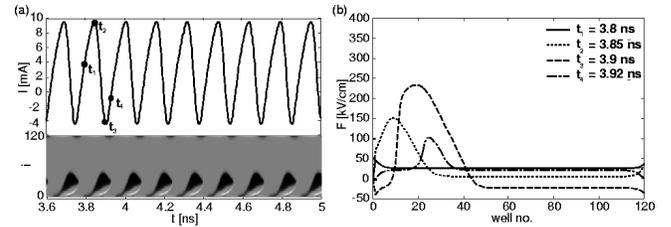}
\caption[a]{Operation of the SL device with an
inductor of 10 nH and a resistor of 25 $\Omega$ in series
($U_{dc}=2$V, $U_{\rm ac}=0$)
(a) Current versus time and space-time plot of 
the electron  density in the SL (as in Fig.1b).  
(b)  Electric field versus position for different times (as indicated in a).
}
\label{FigInductor}
\end{figure}

Fig.~\ref{FigDetail}b
shows the amplitude and the monotonically increasing phase
of the impedance over the whole locking interval.
We extract an impedance 
$Z\approx -i 150 \Omega$ for $\nu_0=2.1 GHz$. Now we compensate 
this impedance by an external circuit with an inductor of
$L=10$ nH and a resistor of 25 $\Omega$ in series with the SL.
(The resistor may result both from connecting wires or due
to radiation damping in a real device.) 
Fig.~\ref{FigInductor} shows that the oscillation mode is
strongly affected: The frequency is increased to
$\nu=5.6$ GHz, the oscillation is more sinusoidal, and
the current amplitude increases. Thus the device 
performance of the SL is strongly improved.
This effect
is due to a different oscillation mode, where the domains
are quenched such that they only traverse a small part of the SL.
As the domain velocity is almost constant, the frequency increases.

We have shown that the amplitude and the phase of  the complex impedance $Z$
exhibit strong variations with the driving frequency $\nu_1$.
The distinct minima of the amplitude of the impedance correspond 
directly to the
frequency locking intervals. In these regions the frequency components of the
current at $\nu_1$ are particularly large, leading to small $|Z|$.
The phase
of the impedance is monotonically increasing with frequency from $-\pi$ to
zero in the locking intervals, which constitutes
a phase shift of $\pi$ compared to the standard
behavior of a driven oscillator. These effects are 
caused by phase synchronization of the
spatio-temporal dynamics of the traveling field domains
in the NDC regime.
Compensating the imaginary part of the impedance in an 
external circuit strongly improves the device performance of 
the SL.

Partial support from Sfb 555
is acknowledged.

\end{multicols}

\end{document}